\documentclass[a4paper,10pt]{article}
\usepackage[utf8]{inputenc}

\usepackage{mathrsfs}
\usepackage{mathtools}

\usepackage{amsmath,amssymb}
\usepackage{biblatex}
\newtheorem{theorem}{Theorem}[section]
\title{Newell-Whitehead-Segel equation,\\
A Simpler Proof}
\author{Luisiana X. Cundin}
\date{January 13, 2026}

\providecommand{\keywords}[1]
{
  \small
  \textbf{\textit{Keywords---}} #1
}

\begin{document}

\maketitle

\begin{abstract}
Previous analysis of the Newell-Whitehead-Segel equation proved the best solution is null; although, the method of solution generated complex nested integrals, therefore, difficult to analyze \cite{NWSgen,NWS2020}. Recent insights into the properties of the convolution integral enable considerable simplification of the solution in the codomain, producing much simpler representations. The inverse Fourier transform of the spectral solution proves to be a non-bijective, null solution, therefore, confirming previous suspicions. Alternative representations of the solution, either expansions or Fujita type solutions, all prove the solution to be a null function.
\end{abstract}


\section{Introduction}
A dogged investigation can reveal new insights into mathematical properties, where previous perspectives are hampered by complexity and lack permitting easy and simple analysis \cite{NWSgen, NWS2020}. Previous investigation of the Newell-Whitehead-Segel (NWS) equation resulted in an infinitly nested solution involving convolutions and time integrals with no simple means of direct analysis; fortunately, a particular property of convolutions enables transferring an exponent into the convolution and onto one or the other function involved in the convolution, leading to a much simpler analytic result in the codomain.

Nonlinear equations often demand multi-valued solutions, of course, the exact nature and breath of multiplicity demanded is equation specific. For the present equation under review, the NWS equation demands a linear Laplacian with a linear, bounded medium response, and an additional, simultaneous nonlinear response, which is typically signed to be unbounded in time; therefore, the solution\textemdash whatever it may be\textemdash is expected to produce two differing results for each specific mapping. Multiplicity raises issues typically not confronted in linear, simply connected spaces, where bijectivity ensures unique, analytic mappings.

In the present analysis, the NWS equation produces a spectral representation with a non-bijective inverse Fourier transform, where the mapping of the entire complex space ($\mathbb{C}$) is null in the spatial domain.

Great care is required in analyzing nonlinear equations, for they present a frustrating, often a circuitous, evasive pathological path; moreover, any linearization employed in finding a solution, for example, expansions, such as, Taylor or perturbation expansions; converting continuous derivatives to discrete representations, such as, Finite Difference or other numerical techniques; \&c, can form representations missing critical properties, such as, poles or multi-valued representations, thus, leading to false, misleading solutions, which are far from faithful to the original equation under investigation. The risk or jeopardy for generally accepted techniques, results, methods of analysis, historical beliefs, \&c, to be erroneous is an ever present danger, despite the width and breath of consensus in a particular neighborhood of a scientific community (consider Phlogiston); because, group-think, bias and falsehoods are certainly always an ever-present threat within any human endeavor.

\section{A simpler solution}
Consider the Newell-Whitehead-Segel equation with arbitrary positive integer power ($n$), \emph{viz.}:
\begin{equation}\label{main}
 \frac{\partial}{\partial t}u(x,t)-\nu\frac{\partial^2}{\partial x^2}u(x,t)+\alpha u(x,t)-\epsilon u(x,t)^n=0
\end{equation}

By inspection, the linear medium response is bounded in time, for the solution will contain an exponentiated time dependency with a negative leading coefficient ($\alpha$). On the other hand, the nonlinear response is typically signed oppositely and is essentially unbounded; albeit, the power of nonlinearity\textemdash being an exponent\textemdash generates Bernoulli type equations (after substituting a convolution for the unknown function), which are rational functions or reciprocal transform or inversion representations. Bernoulli type solutions do give hope for a bounded solution for the nonlinear response, because Bernoulli solutions are reciprocal, bounded solutions, in general.

To solve the system, a convolution integral is substituted for the unknown function, i.e. $u(x,t)=\left(f(x,t)\ast G(x,t)e^{\alpha t}\right)(x)$, where the spatial variable is explicitly stated as the variable involved under the convolution, the convolution is comprised of the linear Green's solution ($Ge^{-\alpha t}$); secondly, an unknown function ($f$) is included in the convolution to account for the nonlinearity in the system. Now, for brevity, all functions will be simply referred to without explicitly stating their dependence upon time and space, e.g. $u=\left(f\ast Ge^{-\alpha t}\right)$.

The derivative with respect to time distributes through the convolution to both functions, but the second order spatial derivative can act singly upon one function involved in the convolution; thus, let's choose the obvious choice of applying the spatial derivative upon Green's solution, i.e. $Ge^{-\alpha t}$. Three terms are generated after applying the partial derivative with respect to time, \emph{viz.}:
\begin{equation}\label{time}
 \left(f'\ast Ge^{-\alpha t}\right)+\left(f\ast G'e^{-\alpha t}\right)-\alpha\left(f\ast Ge^{-\alpha t}\right),
\end{equation}
\noindent where prime ($'$) is used to indicate a time derivative.

The third term in equation (\ref{time}) cancels the third term (linear medium response) in the original partial differential equation (\ref{main}). The second term in equation (\ref{time}) cancels the second-order spatial derivative of Green's function, thus, all linear terms are canceled, explicitly:
\begin{equation}
 \left(f\ast G'e^{-\alpha t}\right)-\nu\left(f\ast \frac{\partial^2}{\partial x^2}Ge^{-\alpha t}\right)=0
\end{equation}

Two terms remain after removing all linear terms, namely, the time derivative of the unknown function ($f$) and the nonlinear term (nonlinear medium response), \emph{ergo},
\begin{equation}\label{rem}
 \left(f'\ast Ge^{-\alpha t}\right)=\epsilon\left(f\ast Ge^{-\alpha t}\right)^n,
\end{equation}
\noindent thus, the problem is dramatically reduced down to the real crux of the problem, i.e. nonlinearity.

It is at this juncture that caused me such consternation in my earlier analysis, because the remaining convolution raised to a power presents a particular difficulty in further analysis. Previously, a sequence of approximations to the exponentiated convolution were made, leading to an iterative solution of Bernoulli equations\ldots ultimately, the resulting solution becomes intractable and not suitable for easy analysis. Fortunately, through a dogged investigation, it was realized the exponent can be moved into the convolution and onto one or the other function involved, and, for a lack of a better name, has been dubbed ''the exponent property for convolution integrals,'' outlined in Theorem \ref{conv}, detailed in the Appendix. With this additional property for convolutions: a more direct, simpler solution is attainable in the codomain, one avoiding iterative integrals \cite{NWS2020}. So, applying the forward Fourier transform to the remainder equation (\ref{rem}), yields:
\begin{equation}
 F'ge^{-\alpha t}=\gamma F^ne^{-\alpha n t}\left(g\ast\ldots_{n-1}\ast g\right)
\end{equation}

The resulting first-order ordinary differential equation with time as a dependent variable proves easy to solve, once it is recognized as a Bernoulli type ordinary differential equation, \emph{viz.}:
\begin{equation}\label{sol}
 F=\left((-n+1)\gamma\int{e^{-\alpha (n-1) t}g^{-1}\left(g\ast\ldots_{n-1}\ast g\right)dt}+1\right)^{\frac{1}{-n+1}}
\end{equation}

The reciprocal of the transformed Green's function ($g^{-1}$) is multiplied by a serial convolution involving transformed Green's functions, plus, an exponentially decaying term, which is raised to a specific power. The solution is complete, where the derivative of the solution ($F$) with respect to time yields a term equal to the last remaining term from the NWS equation, that is to say, equation (\ref{rem}) is solved exactly and generally.

\section{Resolving the Nonlinear Sol'n}
The problem has been reduced to a Bernoulli type solution comprised of a serial convolution of transformed Green's solutions and not an infinitely nested solution of convolutions and time integrations \cite{NWSgen}; actually, the reduced solution does gives hope for an in-depth, piercing analysis.

Several paths do exist analyzing the remainder system, equation (\ref{rem}), where each path yields particular, unique insights into the topological behavior of this type of nonlinear system. The first means of attack is to go ahead and evaluate the convolution of the transformed Green's functions, divide by the reciprocal of the same function, finally, evaluate the time integral. Now, for general exponent of nonlinearity ($n$), the resulting serial convolution simply repeats, which is not prohibitive, in any sense, of solving the resulting serial convolution, but it is repetitive and does not generate any greater insight into the nature of the general solution; for the sake of simplicity, let's consider the exponent of nonlinearity equal to two and focus attention on what is happening in the frequency domain, what restrictions or constraints arise, thus, concentrating our attention on the fundamental nature of this type of system.

Working within the codomain, the serial convolution is evaluated, divided and integrated with respect to time, \emph{viz.}:
\begin{equation}
 F=\frac{1}{\epsilon\int{\left(g\ast g\right) g^{-1} dt}+1}=\frac{1}{\epsilon  \frac{\operatorname{erf}\left(\sqrt{t \left(\alpha -2 \pi ^2
   \nu  s^2\right)}\right)}{2 \sqrt{2} \sqrt{\nu  \left(\alpha -2
   \pi ^2 \nu  s^2\right)}}+1},\{t|t\in (0,t)\}
\end{equation}

The error function is odd, but the square of the frequency variable is even, hence, division by the algebraic term creates a unique behavior; specifically, for values less than the pole, centered on the ratio between the linear medium coefficient and diffusivity constant ($\pm\sqrt{\alpha}/\sqrt{\pi^2\nu}$), the error function is positive, so is the radical algebraic term dividing it, thus, the function is even and positive along the inside of the branch. Stepping beyond the branch point, the frequency term will be greater than the linear coefficient ($\alpha$), leading to an imaginary value for the error function argument and algebraic term, thus, the Jordan curve stretching from infinity to each branch point are real and even. Summarily, to find the inverse Fourier transform for function ($F$), a contour integration is performed, the direction of integration changes, but the integrand has the same sign, thus, the sum of all contours, formed with appropriate Jordan curves yields each complementary curve oppositely signed, thus, cancelling in the sum. Similarly, integrating around each branch point yields oppositely signed results. Ultimately, the resulting inverse Fourier transform is null, \emph{viz.}:
\begin{equation}
 {\mathscr{F}^{-1}\left\{\frac{1}{\epsilon\int{\left(g\ast g\right) g^{-1} dt}+1}\right\}}= 0
\end{equation}

The exact solution yields the null solution, regardless of power of nonlinearity or parameter values, where power of nonlinearity simply raises the entire function ($F$) to some arbitrary power and does not remove the null result.

Earlier analysis certainly suggested, at the least, a spatial delta distribution for a possible solution, but a non-zero solution would force setting the leading coefficient for the solution to be equal to zero. The rationale is a linear Green's solution cannot be the solution for the nonlinear partial differential system. With this simpler formulation, the null result is shown to be universal for the solution \cite{NWSgen}.

\subsection*{Alternate approaches}

An alternative approach to solving the resulting nonlinear equation would be to place emphasis upon the known Green's solution ($g$) in the codomain representation and allow a serial convolution to fall upon the unknown function ($F$), \emph{viz.}:
\begin{equation}
 F'=\epsilon g^{n-1}e^{-\alpha (n-1)t}\left(F\ast\ldots_{n-1}\ast F\right)
\end{equation}

The representation just derived is akin to a convolution type integral often approached by researchers, for example, Fujita type equations and solutions \cite{Fujita}. The resulting Fujita type equation has zero as a potential solution; albeit, separation of variables is most often assumed to generate a non-zero solution, e.g.
\begin{equation}
 h'(t)=\frac{\epsilon}{4\sqrt{\pi\nu}}\frac{e^{-\alpha t}}{\sqrt{t}}h(t)^2
\end{equation}

Obviously, separating the function removes any poles from the solution, thereby, generating non-zero solutions; but, this is obviously incorrect, given the exact solution discussed above. In other words, the resulting approximation forms a time dependent constant; given the solution is equal to the product of the function ($F$) by Green's function, \emph{inept}, $Fge^{-\alpha t}$, and the inverse Fourier transform is essentially the linear solution, therefore, the solution generates a delta function convolved with the linear Green's function in the original space. Since the a linear solution is surely not a sufficient solution for the nonlinear equation, the leading coefficient must be set to zero, which gives the null solution once again.

The method of Separation of Variables (MOSV) has never been validated, yet is universally accepted and used. I, the author, have had reservations regarding the MOSV for some time now, and, have verified the legitimacy of the method in the case of linear partial differential equations, specifically, treating the constant of separation as continuous, the exact Green's heat kernel can be retrieved, after integrating the entire range of values for the constant of separation. The discrete Fourier series (DFT) representation is equal to the continuous Fourier representation, even though, the DFT does require an infinite series to be exact.

Topologically speaking, separating a manifold removes interdependence for all variables separated, thus, tearing the manifold results in pieces singly dependent on one variable; in other words, the dynamic nature of the original space is lost after separating the space into independent, disjointed spaces. For linear manifolds, separating a space and glueing together again does in fact work, but serious reservations exist for partial differential equations with variable coefficients, especially, if a pole is present, such as, the inverse potential ($1/x$). As for nonlinear partial differential equations, the method of Separation of Variables poses tremendous risk, and, in this author's opinion, should be considered invalid in general.

Another approach to solving would be expansions, for example, converting the reciprocal function ($f$), equation (\ref{sol}), into a Neumann series (expansion), which is essentially a geometric series representation. Careful inspection of the criteria for convergence for the geometric series proves the argument must be strictly less than absolute unity. The function is indeterminate at the branch point and cannot be removed, hence, conversion to a geometric series representation is not valid. Regardless, if the argument is converted to a power series, the inverse Fourier transform of each term converts to a serial convolution, and, the inverse Fourier transform of the argument is a null function (the residue theorem is invalid for functions possessing a branch point, thus, a complete Bromwich integral is required), nullity is due to the symmetry of the argument. A Neumann expansion suffers similar constraints to Fujita type solutions, where if a researcher ignores certain conditions, a solution can be generated, which seemingly produces non-zero results, yet, typically, these solutions admit unbounded modes for an index less than two. The fact unbounded terms result in a series expansion should give serious pause for the representation, moreover, it is wholly invalid to cherry-pick lower bounded modes and ignore an infinite set of modes required by summation.

There is lastly one other approach, namely, why not directly move the reciprocal of the transformed Green's function into the convolution, $g^{-1}\left(g\ast g\ldots\right)$, resulting in the division of one or the other transformed Green's function\ldots this simply divides out one or the other occurrence of the transformed Green's function, yielding unity. Now, the convolution of a function over unity yields a definite integral, \emph{viz.}:
\begin{equation}
 \left(1\ast H(s)\right)(s)=\int_{-\infty}^{\infty}{H(s,t)ds}\sim K(t),
\end{equation}
\noindent where the definite integral reduces to some time dependent constant.

Of course, it is a matter to find the inverse Fourier transform of the newly found solution, but the structure of the function ($F$) will, unfortunately, lead to a delta distribution, for the definite integral removes any frequency dependency. \emph{viz.}:
\begin{align}\label{COT}
Fge^{-\alpha t}=& \mathscr{F}^{-1}\left\{\frac{1}{\int{\int_{-\infty}^{\infty}{H(s,t)ds}\ dt}+1}\right\}\ast Ge^{-\alpha t}=\\\label{COT2}
& AK(t)\left(\delta(x)\ast Ge^{-\alpha t}\right)(x)=AK(t)Ge^{-\alpha t},
\end{align}
\noindent where leading coefficient ($A$) is arbitrary and function of time ($K$) accounts for any residual time dependency.

Taking a step back, the above solution essentially reduces to a delta distribution for the function ($f$), which leaves a time dependent linear solution. Ultimately, regardless of the amplitude, the leading coefficient ($A$) must be set to a constant, preferably, a null solution.

Considering a serial convolution of transformed Green's functions divided by the reciprocal of the same, the initial reduction of the first convolution to a constant rifles through the remaining serial convolution to reduce the entire convolution to a constant, in other words, a very repetitive, similar solution, where the ultimate solution involves something similar to equation (\ref{COT2}). It would be remiss not to mention that no matter the time dependency in-front of the transformed Green's solution in equation (\ref{COT}), the leading coefficient ($A$) is ideally set to null, thus, no matter the degree of freedom considered (spatial dimensions), \&c, the solution is the same, i.e. $u(x,y,z;t)=0$.

\section{Theorems}

\setcounter{equation}{0}
\begin{theorem}[Exponent property for Convolutions]\label{conv}
A consequential property of convolution integrals enables distributing an exponent to either function involved, \emph{viz.}:

\begin{equation}
 (f\ast g)^n=(f\ast g^n)=(f^n\ast g)
\end{equation}

The proof is simple, employing the convolution theorem from Fourier theory converts a convolution raised to an arbitrary positive integer ($n$) into a serial set of convolutions comprised of the product of each original function's Fourier transform, \emph{viz.}:

\begin{align}
 &(f\ast G)^n\subset \left(Fg\ast\ldots  \ast_{n-1}\ast Fg\right)=\label{c1}\\
 &\hspace{12pt} F^n\left(g\ast\ldots\ast_{n-1}\ast g\right)\supset \left(f\ast\ldots_{n-1}\ast f\right)\ast G^n\label{c2},
\end{align}
\noindent where the Fourier transform of a lower case lettered function ($f$) is represented by its capital letter, e.g. $\mathscr{F}\left\{f\right\}=F$, furthermore, Bracewell's convention for symbolizing a forward Fourier transform ($\subset$) and inverse Fourier transform ($\supset$) are conveniently employed for clarity, \emph{ergo}, $f\subset F$; additionally, convolution is represented by an asterisk ($\ast$)\cite{bracewell}.

If the exponent is unity, the standard convolution theorem yields the product of each transformed function, $\left(f\ast G\right)\subset FG $. For powers greater than unity, a series of convolutions over the product of the transformed functions do unfurl, thus, $Fg\ast Fg\ast\ldots$, where the number of convolutions resulting are equal to the power of the exponent minus one, thus, the last convolution would equal the total number of convolutions indicated by the exponent ($n$); this property is demonstrated in equation (\ref{c1}).

It is another property of convolutions that a convolution multiplied by a function distributes arbitrarily into the convolution, that is to say,

\begin{equation}
 h\left(f\ast G\right)=\left(hf\ast G\right)=\left(f\ast hG\right);
\end{equation}
\noindent therefore, removing all transformed functions ($F$) from a serial convolution yields that function raised to the number of occurrences found throughout the serial convolution, as demonstrated in equation (\ref{c2}); and, after applying the inverse Fourier transform, the original function ($G$) is retrieved as a product, and, the original function ($f$) is now represented as a serial convolution. A similar result is generated if the other function ($G$) is successively pulled from the convolution, which leads to similar result, except, now the other function is raised to a power and a serial self-repeating convolution is left for the other function.

Ultimately, there is tremendous freedom and several resultant configurations can be derived from a convolution integral raised to an integer power ($n$).

\end{theorem}

\printbibliography

\end{document}